\pdfoutput=1
\documentclass[12pt,a4paper]{article}

\usepackage{ifthen} 
\newboolean{pdflatex}
\setboolean{pdflatex}{true} 

\newboolean{articletitles}
\setboolean{articletitles}{true} 

\newboolean{uprightparticles}
\setboolean{uprightparticles}{false} 

\def\paperauthors{LHCb collaboration} 
\def\paperasciititle{ Branching fraction measurements of eta and eta prime anomalous decays} 
\def\papertitle{ Branching fraction measurements of \etaz and \etaprpdg anomalous decays} 
\def\paperkeywords{{High Energy Physics}, {LHCb}} 
\def\papercopyright{\the\year\ CERN for the benefit of the LHCb collaboration} 
\def\paperlicence{CC BY 4.0 licence}
\def\paperlicenceurl{https://creativecommons.org/licenses/by/4.0/}

\newif\ifEnableSectionTOCLinks
\EnableSectionTOCLinksfalse 

\usepackage[top=1in, bottom=1.25in, left=1in, right=1in]{geometry}

\usepackage{microtype}
\usepackage{lineno}  
\usepackage{xspace} 
\usepackage{caption} 

\usepackage{booktabs}
\usepackage{tabularx}

\usepackage{subcaption}
\usepackage{array}
\newcolumntype{C}[1]{>{\centering\arraybackslash}p{#1}}

\usepackage{verbatim}

\usepackage{graphicx}  
\usepackage{color}
\usepackage{colortbl}
\graphicspath{{./figs/}} 

\usepackage{amsmath} 
\usepackage{amssymb}
\usepackage{amsfonts}
\usepackage{upgreek} 

\newcommand*\patchAmsMathEnvironmentForLineno[1]{%
\expandafter\let\csname old#1\expandafter\endcsname\csname #1\endcsname
\expandafter\let\csname oldend#1\expandafter\endcsname\csname
end#1\endcsname
 \renewenvironment{#1}%
   {\linenomath\csname old#1\endcsname}%
   {\csname oldend#1\endcsname\endlinenomath}%
}
\newcommand*\patchBothAmsMathEnvironmentsForLineno[1]{%
  \patchAmsMathEnvironmentForLineno{#1}%
  \patchAmsMathEnvironmentForLineno{#1*}%
}
\AtBeginDocument{%
\patchBothAmsMathEnvironmentsForLineno{equation}%
\patchBothAmsMathEnvironmentsForLineno{align}%
\patchBothAmsMathEnvironmentsForLineno{flalign}%
\patchBothAmsMathEnvironmentsForLineno{alignat}%
\patchBothAmsMathEnvironmentsForLineno{gather}%
\patchBothAmsMathEnvironmentsForLineno{multline}%
\patchBothAmsMathEnvironmentsForLineno{eqnarray}%
}

\usepackage[pdftex,
            pdfauthor={\paperauthors},
            pdftitle={\paperasciititle},
            pdfkeywords={\paperkeywords}]{hyperref}
\usepackage{hyperxmp}
\hypersetup{
    pdfcopyright={Copyright (C) \papercopyright},
    pdflicenseurl={\paperlicenceurl}
}

\usepackage[colorinlistoftodos,textsize=scriptsize]{todonotes}

\usepackage[bottom,flushmargin,hang,multiple]{footmisc}

\usepackage[all]{hypcap} 

\usepackage{xspace}
\usepackage{upgreek}

\def\lhcb   {\mbox{LHCb}\xspace}

\def\MagUp {\mbox{\em Mag\kern -0.05em Up}\xspace}

\def\hltone {HLT1\xspace}
\def\hlttwo {HLT2\xspace}

\ifthenelse{\boolean{uprightparticles}}%
{\def\Palpha      {\ensuremath{\upalpha}\xspace}
 
 \def\Pgamma      {\ensuremath{\upgamma}\xspace}

 \def\Peta        {\ensuremath{\upeta}\xspace}

 \def\Pmu         {\ensuremath{\upmu}\xspace}

 \def\Ppi         {\ensuremath{\uppi}\xspace}

 \def\Pphi        {\ensuremath{\upphi}\xspace}

 \def\Ppsi        {\ensuremath{\uppsi}\xspace}

 \def\PDelta      {\ensuremath{\Delta}\xspace}
 \def\PXi         {\ensuremath{\Xi}\xspace}
 \def\PLambda     {\ensuremath{\Lambda}\xspace}
 \def\PSigma      {\ensuremath{\Sigma}\xspace}
 \def\POmega      {\ensuremath{\Omega}\xspace}
 \def\PUpsilon    {\ensuremath{\Upsilon}\xspace}
 \let\oldPi\Pi
 \def\PPi         {\ensuremath{\oldPi}\xspace}

 \def\PB      {\ensuremath{\mathrm{B}}\xspace}
 \def\PD      {\ensuremath{\mathrm{D}}\xspace}
 \def\PJ      {\ensuremath{\mathrm{J}}\xspace}
 \def\PK      {\ensuremath{\mathrm{K}}\xspace}
 \def\Pb      {\ensuremath{\mathrm{b}}\xspace}
 \def\Pc      {\ensuremath{\mathrm{c}}\xspace}
 \def\Pd      {\ensuremath{\mathrm{d}}\xspace}
 \def\Pe      {\ensuremath{\mathrm{e}}\xspace}
 \def\Ps      {\ensuremath{\mathrm{s}}\xspace}
 
 \def\Pu      {\ensuremath{\mathrm{u}}\xspace}
 \def\thebaroffset{0.0em}
}
{\def\Palpha      {\ensuremath{\alpha}\xspace}
 
 \def\Pgamma      {\ensuremath{\gamma}\xspace}

 \def\Peta        {\ensuremath{\eta}\xspace}

 \def\Pmu         {\ensuremath{\mu}\xspace}

 \def\Ppi         {\ensuremath{\pi}\xspace}

 \def\Pphi        {\ensuremath{\phi}\xspace}

 \def\Ppsi        {\ensuremath{\psi}\xspace}
 
 \mathchardef\PDelta="7101
 \mathchardef\PXi="7104
 \mathchardef\PLambda="7103
 \mathchardef\PSigma="7106
 \mathchardef\POmega="710A
 \mathchardef\PUpsilon="7107
 \mathchardef\PPi="7105
 \def\PB      {\ensuremath{B}\xspace}
 \def\PD      {\ensuremath{D}\xspace}
 \def\PJ      {\ensuremath{J}\xspace}
 \def\PK      {\ensuremath{K}\xspace}
 \def\Pb      {\ensuremath{b}\xspace}
 \def\Pc      {\ensuremath{c}\xspace}
 \def\Pd      {\ensuremath{d}\xspace}
 \def\Pe      {\ensuremath{e}\xspace}
 \def\Ps      {\ensuremath{s}\xspace}
 
 \def\Pu      {\ensuremath{u}\xspace}
 \def\thebaroffset{0.18em}
}
\newcommand{\offsetoverline}[2][\thebaroffset]{\kern #1\overline{\kern -#1 #2}}%

\makeatletter
\ifcase \@ptsize \relax
  \newcommand{\miniscule}{\@setfontsize\miniscule{4}{5}}
\or
  \newcommand{\miniscule}{\@setfontsize\miniscule{5}{6}}
\or
  \newcommand{\miniscule}{\@setfontsize\miniscule{5}{6}}
\fi
\makeatother

\DeclareRobustCommand{\optbar}[1]{\shortstack{{\miniscule (\rule[.5ex]{1.25em}{.18mm})}
  \\ [-.7ex] $#1$}}

\def\en         {{\ensuremath{\Pe^-}}\xspace}   
\def\ep         {{\ensuremath{\Pe^+}}\xspace}

\def\mup        {{\ensuremath{\Pmu^+}}\xspace}
\def\mun        {{\ensuremath{\Pmu^-}}\xspace} 

\def\mumu       {{\ensuremath{\Pmu^+\Pmu^-}}\xspace}

\def\g      {{\ensuremath{\Pgamma}}\xspace}

\def\uquark    {{\ensuremath{\Pu}}\xspace}

\def\dquark    {{\ensuremath{\Pd}}\xspace}

\def\squark    {{\ensuremath{\Ps}}\xspace}

\def\cquark    {{\ensuremath{\Pc}}\xspace}

\def\bquark    {{\ensuremath{\Pb}}\xspace}

\def\pion   {{\ensuremath{\Ppi}}\xspace}

\def\pip    {{\ensuremath{\pion^+}}\xspace}
\def\pim    {{\ensuremath{\pion^-}}\xspace}

\def\pipi   {{\ensuremath{\pip\pim}}\xspace}

\def\kaon    {{\ensuremath{\PK}}\xspace}

\def\KorKbar {\kern \thebaroffset\optbar{\kern -\thebaroffset \PK}{}\xspace}

\def\Kp      {{\ensuremath{\kaon^+}}\xspace}
\def\Km      {{\ensuremath{\kaon^-}}\xspace}

\newcommand{\etaz}{\ensuremath{\Peta}\xspace}
\newcommand{\etapr}{\ensuremath{\Peta^{\prime}}\xspace}
\newcommand{\phiz}{\ensuremath{\Pphi}\xspace}

\def\D       {{\ensuremath{\PD}}\xspace}

\def\DorDbar {\kern \thebaroffset\optbar{\kern -\thebaroffset \PD}\xspace}
\def\Dz      {{\ensuremath{\D^0}}\xspace}

\def\Dp      {{\ensuremath{\D^+}}\xspace}
\def\Dm      {{\ensuremath{\D^-}}\xspace}

\def\DpDm    {\ensuremath{\Dp {\kern -0.16em \Dm}}\xspace}

\def\Ds      {{\ensuremath{\D^+_\squark}}\xspace}

\def\DporDsp {{\ensuremath{\D_{(\squark)}^+}}\xspace}

\def\B       {{\ensuremath{\PB}}\xspace}

\def\BorBbar {\kern \thebaroffset\optbar{\kern -\thebaroffset \PB}\xspace}

\def\Bd      {{\ensuremath{\B^0}}\xspace}

\def\BdorBdbar {\kern \thebaroffset\optbar{\kern -\thebaroffset \Bd}\xspace}

\def\Bs      {{\ensuremath{\B^0_\squark}}\xspace}

\def\BsorBsbar {\kern \thebaroffset\optbar{\kern -\thebaroffset \Bs}\xspace}

\def\jpsi     {{\ensuremath{{\PJ\mskip -3mu/\mskip -2mu\Ppsi}}}\xspace}

\def\Y#1S{\ensuremath{\PUpsilon{(#1S)}}\xspace}

\def\LorLbar     {\kern \thebaroffset\optbar{\kern -\thebaroffset \PLambda}\xspace}

\def\BF         {{\ensuremath{\mathcal{B}}}\xspace}
\def\BR         {\BF}

\newcommand{\decay}[2]{\mbox{\ensuremath{#1\!\to #2}}\xspace}

\def\to                 {\ensuremath{\rightarrow}\xspace}

\def\grpsuthree {{\ensuremath{\mathrm{SU}(3)}}\xspace}

\def\AT#1     {\ensuremath{A_{\mathrm{T}}^{#1}}\xspace}           

\def\C#1      {\ensuremath{\mathcal{C}_{#1}}\xspace}                       
\def\Cp#1     {\ensuremath{\mathcal{C}_{#1}^{'}}\xspace}                    
\def\Ceff#1   {\ensuremath{\mathcal{C}_{#1}^{\mathrm{(eff)}}}\xspace}        
\def\Cpeff#1  {\ensuremath{\mathcal{C}_{#1}^{'\mathrm{(eff)}}}\xspace}       
\def\Ope#1    {\ensuremath{\mathcal{O}_{#1}}\xspace}                       
\def\Opep#1   {\ensuremath{\mathcal{O}_{#1}^{'}}\xspace}                    

\def\dkpicf     {\decay{\Dz}{\Km\pip}}

\newcommand{\aunit}[1]{\ensuremath{\text{\,#1}}}

\newcommand{\tev}{\aunit{Te\kern -0.1em V}\xspace}
\newcommand{\gev}{\aunit{Ge\kern -0.1em V}\xspace}
\newcommand{\mev}{\aunit{Me\kern -0.1em V}\xspace}
\newcommand{\kev}{\aunit{ke\kern -0.1em V}\xspace}
\newcommand{\ev}{\aunit{e\kern -0.1em V}\xspace}

\newcommand{\mevc}{\ensuremath{\aunit{Me\kern -0.1em V\!/}c}\xspace}
\newcommand{\gevc}{\ensuremath{\aunit{Ge\kern -0.1em V\!/}c}\xspace}
\newcommand{\mevcc}{\ensuremath{\aunit{Me\kern -0.1em V\!/}c^2}\xspace}
\newcommand{\gevcc}{\ensuremath{\aunit{Ge\kern -0.1em V\!/}c^2}\xspace}

\def\fb   {\ensuremath{\aunit{fb}}\xspace}
\def\invfb   {\ensuremath{\fb^{-1}}\xspace}

\newcommand{\stat}{\aunit{(stat)}\xspace}
\newcommand{\syst}{\aunit{(syst)}\xspace}
\newcommand{\ext}{\aunit{(ext)}\xspace}

\def\gsim{{~\raise.15em\hbox{$>$}\kern-.85em
          \lower.35em\hbox{$\sim$}~}\xspace}
\def\lsim{{~\raise.15em\hbox{$<$}\kern-.85em
          \lower.35em\hbox{$\sim$}~}\xspace}

\def\PDF {PDF\xspace}

\def\sqs   {\ensuremath{\protect\sqrt{s}}\xspace}

\def\pt         {\ensuremath{p_{\mathrm{T}}}\xspace}

\def\evtgen     {\mbox{\textsc{EvtGen}}\xspace}

\def\geant      {\mbox{\textsc{Geant4}}\xspace}

\def\photos     {\mbox{\textsc{Photos}}\xspace}

\def\pythia     {\mbox{\textsc{Pythia}}\xspace}

\def\tell1  {TELL1\xspace}
\def\ukl1   {UKL1\xspace}

\newcommand{\phz}{\phantom{0}}
\newcommand{\lhcborcid}[1]{\href{https://orcid.org/#1}{\hspace*{0.1em}\raisebox{-0.45ex}{\includegraphics[width=1em]{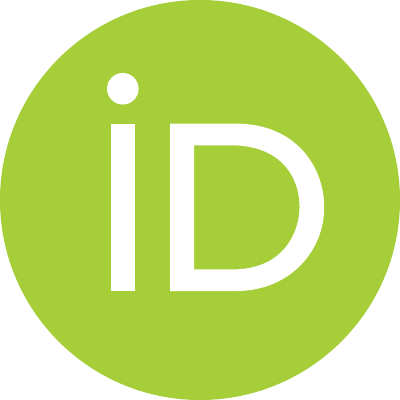}}}}

\newcommand{\BRof}[1]{\ensuremath{\BF(#1)}\xspace}

\newcommand{\etaprpdg}{\ensuremath{\etapr(958)}\xspace}
\newcommand{\etaoretap}{\ensuremath{\eta^{(\prime)}}\xspace}
\newcommand{\gmtwo}{\ensuremath{(g - 2)_\mu}\xspace}

\newcommand{\etafb}{\ensuremath{\eta \to \pip\pim \mup \mun}\xspace}
\newcommand{\etapfb}{\ensuremath{\etapr \to \pip\pim \mup \mun}\xspace}
\newcommand{\etaptb}{\ensuremath{\etapr \to \mup \mun}\xspace}
\newcommand{\etappdgfb}{\ensuremath{\etaprpdg \to \pip\pim \mup \mun}\xspace}
\newcommand{\etappdgtb}{\ensuremath{\etaprpdg \to \mup \mun}\xspace}
\newcommand{\etamm}{\ensuremath{\etaz \to \mup \mun}\xspace}

\newcommand{\phimm}{\ensuremath{\phiz \to \mup \mun}\xspace}

\newcommand{\dsphi}{\ensuremath{\Ds \to \phiz \pip }\xspace}

\newcommand{\dsphifb}{\ensuremath{\Ds \to \phiz(\to \mup \mun) \pip \pim \pip }\xspace}

\newcommand{\dphipi}{\ensuremath{\DporDsp \to \phiz \pip}\xspace}
\newcommand{\dphimmpi}{\ensuremath{\DporDsp \to \phiz(\to\mup\mun) \pip}\xspace}

\newcommand{\detapi}{\ensuremath{\DporDsp \to \etaoretap \pip}\xspace}

\newcommand{\figref}[1]{Fig.~\ref{#1}}

\newcommand{\tabref}[1]{Table~\ref{#1}}

\newcommand{\refeq}[1]{Eq.~(\ref{#1})}

\hypersetup{
  colorlinks   = true, 
  urlcolor     = blue, 
  linkcolor    = blue, 
  citecolor    = red   
}

\ifEnableSectionTOCLinks
    \usepackage[explicit]{titlesec} 
    
    \let\oldcontentsline\contentsline
    \renewcommand

    \titleformat{\section}{\normalfont\Large\bf}{\hyperlink{tocsection.\thesection}{{\thesection} \parbox[t]{\dimexpr\textwidth-1pc}{#1}}}{1pc}{}

    \titleformat{\subsection}{\normalfont\bf}{\hyperlink{tocsubsection.\thesubsection}{{\thesubsection} \parbox[t]{\dimexpr\textwidth-1pc}{#1}}}{1pc}{}

    \titleformat{name=\section,numberless}[display]{}{}{0pt}{\normalfont\Huge\bfseries #1}
\fi

\usepackage{cite} 
\usepackage{mciteplus}

\usepackage{longtable} 

\begin{document}

\renewcommand{\thefootnote}{\fnsymbol{footnote}}
\setcounter{footnote}{1}


\begin{titlepage}
\pagenumbering{roman}

\vspace*{-1.5cm}
\centerline{\large EUROPEAN ORGANIZATION FOR NUCLEAR RESEARCH (CERN)}
\vspace*{1.5cm}
\noindent
\begin{tabular*}{\linewidth}{lc@{\extracolsep{\fill}}r@{\extracolsep{0pt}}}
\ifthenelse{\boolean{pdflatex}}
{\vspace*{-1.5cm}\mbox{\!\!\!\includegraphics[width=.14\textwidth]{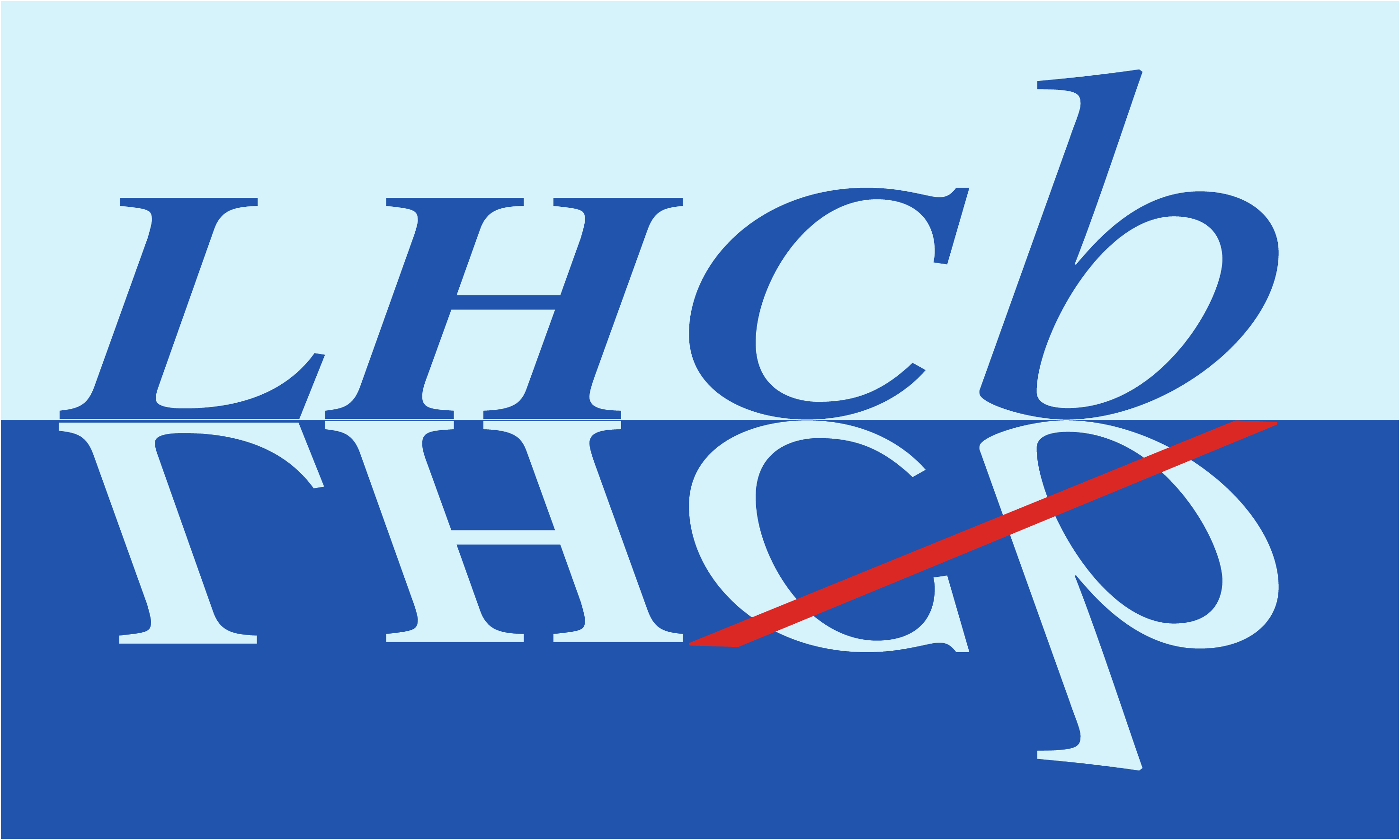}} & &}%
{\vspace*{-1.2cm}\mbox{\!\!\!\includegraphics[width=.12\textwidth]{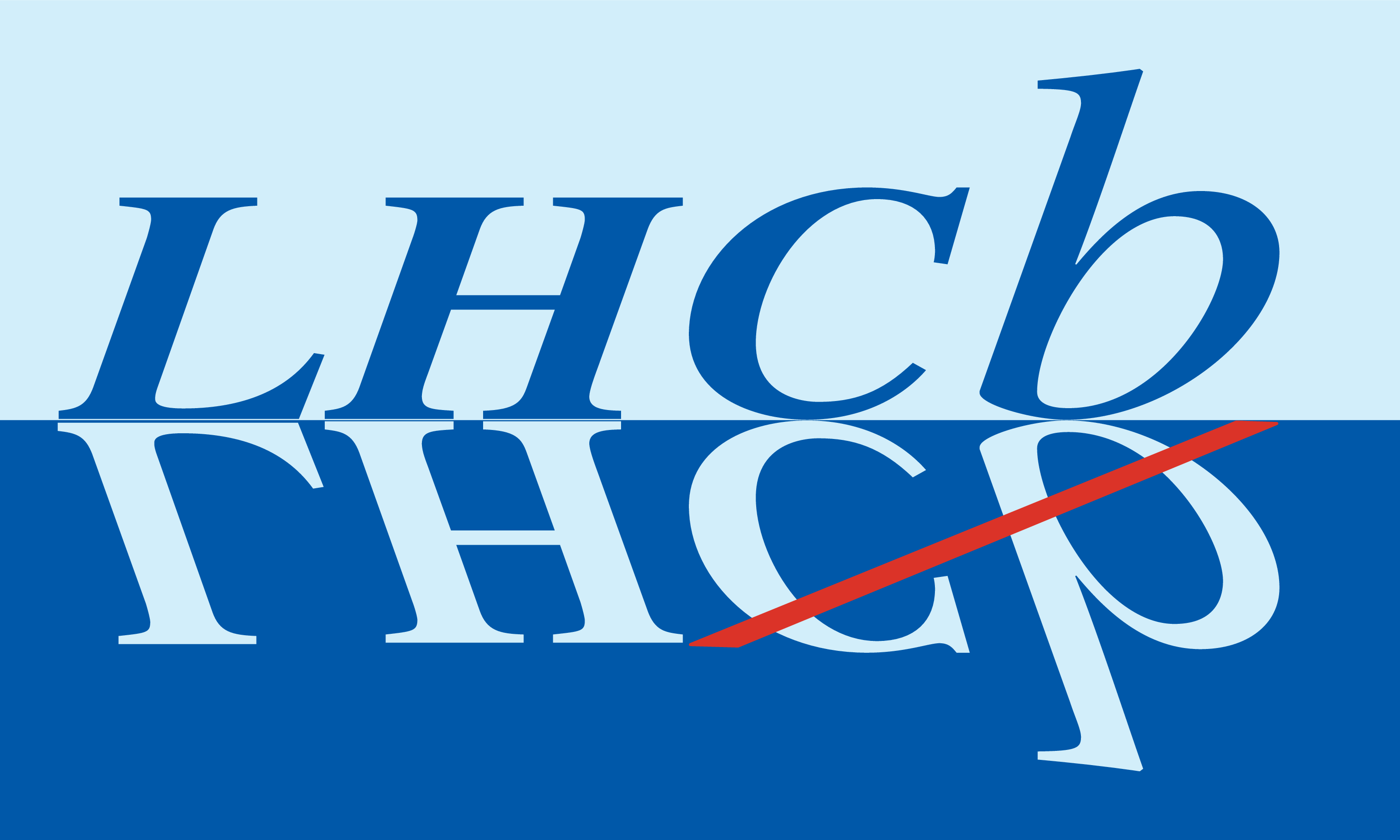}} & &}%
\\
 & & CERN-EP-2026-238 \\  
 & & LHCb-PAPER-2025-047 \\  
 & & September 4, 2026 \\ 
 & & \\
\end{tabular*}

\vspace*{4.0cm}

{\normalfont\bfseries\boldmath\huge
\begin{center}
  \papertitle 
\end{center}
}

\vspace*{2.0cm}

\begin{center}
\paperauthors\footnote{Authors are listed at the end of this Letter.}
\end{center}

\vspace{\fill}

\begin{abstract}
  \noindent
  Branching fractions of $\eta$ and $\eta'(958)$ anomalous decays are measured using $pp$ collision data at a center-of-mass energy of $\sqrt{s}=13\,\text{TeV}$ collected by the LHCb experiment and
  corresponding to an integrated luminosity of $5.4\,\text{fb}^{-1}$.
   The measured values and upper limits are
  \begin{align*}
      \mathcal{B}(\eta\to\mu^+\mu^-)  &= (5.02\pm0.14\pm0.18\pm0.26)\times10^{-6} , \\
      \mathcal{B}(\eta'(958)\to\mu^+\mu^-) &<1.9\times10^{-7}\textrm{ at 90\% CL} , \\
      \mathcal{B}(\eta\to\pi^+\pi^-\mu^+\mu^-) &<5.4\times10^{-7}\textrm{ at 90\% CL}, \\
      \mathcal{B}(\eta'(958)\to\pi^+\pi^-\mu^+\mu^-)  &= (2.73\pm0.14\pm0.13\pm0.14)\times10^{-5} ,
  \end{align*} 
  where the uncertainties are, respectively, statistical, systematic,
  and due to the external branching fraction of the 
  $\phi(1020)\to\mu^+\mu^-$ decay, which is used for normalization.
  This marks the first experimental search for the $\eta'(958)$ decay, 
  and the most precise measurement of the branching fraction $\mathcal{B}(\eta\to\mu^+\mu^-)$ to date.
\end{abstract}

\vspace*{2.0cm}

\begin{center}
  Submitted to
  Phys.~Rev.~Lett. 
\end{center}

\vspace{\fill}

{\footnotesize 
\centerline{\copyright~\papercopyright. \href{\paperlicenceurl}{\paperlicence}.}}
\vspace*{2mm}

\end{titlepage}


\newpage
\setcounter{page}{2}
\mbox{~}


\renewcommand{\thefootnote}{\arabic{footnote}}
\setcounter{footnote}{0}

\cleardoublepage


\pagestyle{plain} 
\setcounter{page}{1}
\pagenumbering{arabic}



The $\etaz$--$\etaprpdg$ system provides an ideal laboratory to test the regime of validity
of effective field theories (EFTs) of Quantum Chromodynamics (QCD) in the low-energy range. 
One example is chiral perturbation theory (ChPT)~\cite{Pich:1995bw,Ecker:1994gg,Bernard:2006gx,Bijnens:2006zp},
which assumes massless \uquark, \dquark, and \squark quarks.
Under this assumption, the QCD Lagrangian is spontaneously broken at energy scales of $\sim1\gev$, 
giving rise to 8 Goldstone bosons, known as the pseudoscalar meson octet. 
This includes the $\eta_8$ state, which is identified as the physical \etaz meson 
under exact \grpsuthree symmetry~\cite{Weinberg:1996kr}. 
The heavier \etaprpdg meson, hereafter referred to as~\etapr, arises from the breaking of the axial $U(1)_A$ symmetry.

\begin{figure}[b]
\centering
    \begin{subfigure}{0.48\linewidth}
        \includegraphics[width=1\linewidth]{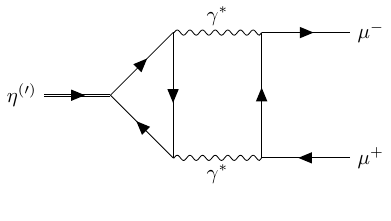}
    \end{subfigure}
    \begin{subfigure}{0.48\linewidth}
        \includegraphics[width=1\linewidth]{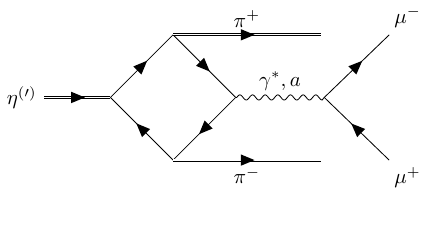}
    \end{subfigure}
\caption{Feynman diagrams representing the (left) triangle anomaly and (right) box anomaly,
which contribute to the decays of $\eta^{(\prime)}$ mesons to $\mu^+\mu^-$ and $\pi^+\pi^-\mu^+\mu^-$, respectively.
In certain BSM models, the $\mu^+\mu^-$ decay in the box anomaly may be mediated not only by virtual photons~ $\gamma^\ast$ but also by a dark photon or an axion-like particle, collectively denoted $a$~\cite{Reece:2009un,Aloni:2018vki,Landini:2019eck}.
}
\label{fig:feynman}
\end{figure}

Rare \etaoretap decays, such as those into the \mumu and \mbox{\pipi\mumu} final states, 
are particularly sensitive probes of theoretical predictions by EFTs,
as small higher-order corrections to the decay rates can have a large relative effect.
Decays of these mesons into dimuon pairs proceed via the triangle anomaly, 
which involves an intermediate two-photon state (\figref{fig:feynman}, left). 
This coupling of a pseudoscalar to two photons is forbidden by the parity structure 
of the conventional chiral Lagrangian, and only arises because chiral symmetry is anomalously broken at the quark level. 
The amplitude of the process is expressed in terms of the pseudoscalar-photon transition form factors (TFFs)~\cite{Masjuan:2015cjl}. 
Decays into two pions and two muons are induced by the box anomaly (\figref{fig:feynman}, right), 
the analogous parity-violating coupling of a pseudoscalar to two pions and a photon, 
and can be described within different EFTs, such as the hidden gauge model~\cite{Petri:2010ea}, 
the chiral unitary approach~\cite{Borasoy:2007dw}, or through the vector-meson dominance (VMD) model~\cite{Faessler:1999de}.

Furthermore, all of these decays are sensitive to both beyond the Standard Model~(BSM) axial and pseudoscalar contributions~\cite{Masjuan:2015cjl,Davoudiasl:2012ag,Davoudiasl:2012qa,Chang:2008np,Carlson:2012pc,Karshenboim:2014tka}.
In particular, the four-body decays could also receive BSM contributions from $\etapr \to \pipi a$ decays, 
where $a$ stands for a dark photon~\cite{Reece:2009un} or an axion-like particle~\cite{Aloni:2018vki,Landini:2019eck}, 
decaying to a pair of muons. 

Further motivation for studying anomalous \etaoretap decays arises from the muon magnetic moment puzzle, \gmtwo~\cite{Aoyama:2020ynm,Aliberti:2025beg}. 
Hadronic contributions to \gmtwo depend on ChPT TFFs, 
which parameterize pseudoscalar-to-two-photon couplings. 
Precision on these TFFs can be substantially improved by measuring \etaz and \etapr branching fractions 
(\BR)~\cite{Huong:2016gob,Knecht:1999gb,RamseyMusolf:2002cy,Gan:2020aco}.

This Letter presents the study of the \etaz and \etapr decays
to \mumu and \mbox{\pipi\mumu} final states, 
using proton-proton ($pp$) collision data corresponding to an integrated luminosity of \mbox{$5.4\invfb$} 
collected with the LHCb detector between 2016 and 2018, at a center-of-mass energy of \mbox{$\sqs=13\tev$}.
A search for BSM contributions mediated by an axion-like particle or a dark photon is also presented in this Letter.
Theoretical predictions for all four signal modes, 
as well as the state-of-the-art experimental results, are provided in~\tabref{tab:status}.
The four-order-of-magnitude difference between the \mbox{\etafb} and \mbox{\etapfb} decays arises from
the smaller mass of the \etaz meson compared to the \etapr,
which limits the available phase space for the decay products.
The BESIII collaboration recently reported updated studies
of the \mbox{\etamm}~\cite{BESIII:2025jft}, \mbox{\etafb}~\cite{BESIII:2025cky}, 
\mbox{\etapfb}~\cite{BESIII:2024awu}, and \mbox{$\eta\to \ep\en\ep\en$}~\cite{BESIII:2026izr} decays. 
The CMS experiment has also recently studied various rare $\eta$ decays, 
reporting observations of \mbox{$\eta\to\mumu\mumu$}~\cite{CMS:2023thf} and $\eta\to\mumu \ep\en$~\cite{CMS:2026qbd}. 
The branching fraction of~\etaptb has never been measured.

\begin{table}[t]
    \centering
    \caption{Theoretical predictions for various models and best measurements of the branching fractions 
    of the decays of interest in this analysis.}
    \label{tab:status}
    \resizebox{\linewidth}{!}{
    \setlength{\tabcolsep}{3pt}
    \begin{tabular}{ll|rcc|c} 
    \hline
    Decay mode & Unit & \multicolumn{3}{c|}{Theory} & Experiment \\
    \hline
    &  & \multicolumn{1}{c}{ChPT} & VMD & Hidden gauge &  \\
    \hline
    \etamm   & $10^{-6}$ & $4.72^{+0.05}_{-0.21}$~\cite{Masjuan:2015cjl} & --- & --- & $\phantom{000}5.8  \pm 0.8\phantom{I\,}$~\cite{PDG2024} \\ 
    \etaptb   & $10^{-7}$ & $1.36^{+0.29}_{-0.26}$~\cite{Masjuan:2015cjl} & --- & --- & --- \\  
    \etafb    & $10^{-9}$ & $\phantom{0}7.5\phantom{0}^{+4.5\phantom{0}}_{-2.7\phantom{0}}$~\cite{Borasoy:2007dw}      & $8.64\pm0.25$~\cite{Faessler:1999de} & $8.65\pm0.39$~\cite{Petri:2010ea} & $<400~(90\%\,\mathrm{CL})$~\cite{BESIII:2025cky} \\
    \etapfb   & $10^{-5}$ & $1.57^{+0.96}_{-0.75}$~\cite{Borasoy:2007dw}  & $2.41\pm0.25$~\cite{Faessler:1999de} & $2.20\pm0.30$~\cite{Petri:2010ea} & $\phantom{00}2.13\pm0.13 \phantom{\,}$~\cite{PDG2024} \\ 
    \hline 
    \end{tabular}
    }
\end{table}

The LHCb detector~\cite{LHCb-DP-2008-001,LHCb-DP-2014-002} is a single-arm forward spectrometer covering the pseudorapidity range 
$2 < \eta < 5$, designed for the study of particles containing \bquark or \cquark quarks.
The detector elements that are particularly
relevant to this analysis include: a silicon-strip vertex detector surrounding the $pp$ interaction
region that allows \cquark\ and \bquark\ hadrons to be identified from their characteristically long
flight distance; a tracking system that provides a measurement of the momentum, $p$, of charged
particles; and two ring-imaging Cherenkov detectors that are able to discriminate between
different species of charged hadrons.

Events are first filtered by a hardware trigger, known as L0~\cite{LHCb-DP-2012-004}, which uses information from the calorimeter and the muon system to select final-state particles with \pt above a few\gevc. Subsequently,
a full event reconstruction is applied in the high-level software trigger (HLT), which runs in two separate steps (HLT1 and HLT2). 

Simulation is used to model the effects of the detector acceptance and
selection requirements. 
In the simulation, $pp$ collisions are generated using
\pythia8~\cite{Sjostrand:2007gs} 
with a specific \lhcb configuration~\cite{LHCb-PROC-2010-056}.
Decays of unstable particles are described by \evtgen~\cite{Lange:2001uf}, 
in which final-state radiation is generated using \photos~\cite{davidson2015photos}.
The interaction of the generated particles with the detector, and its response,
are implemented using the \geant toolkit~\cite{Allison:2006ve,Agostinelli:2002hh} 
as described in Ref.~\cite{LHCb-PROC-2011-006}.

A highly effective way to study \etaoretap decays at \lhcb 
is via their abundant production in charm meson decays~\cite{Huong:2016gob}.
In particular, the $D_{(s)}^+\to\etaoretap\pip$ decays, 
with branching fractions between $0.4\%$ and $4\%$, provide 
large and high-purity samples of \etaz and \etapr mesons.
In the same regard, \dphimmpi decays provide excellent normalization modes, 
owing to their comparatively large branching fraction and final-state topology closely matching that of the signal channels.
Charge conjugation is implied throughout this Letter.

The branching fraction of each signal mode is determined with respect to the corresponding normalization channel \mbox{\dphimmpi}, 
using the signal yield ($N_{\textrm{sig}}$) and the normalization factor ($\alpha$), as
\begin{equation}
    \BR(\decay{\eta^{(\prime)}}{(\pip\pim)\mup\mun}) = \alpha \times N_{\text{sig}},
\end{equation}
where 
\begin{equation}
\label{eq:alpha}
    \alpha = \frac{1}{N_{\text{norm}}} \times \frac{\BRof\dphipi\times\BRof\phimm}{\BRof\detapi}\times\frac{\varepsilon_{\text{norm}}}{\varepsilon_{\text{sig}}}.
\end{equation}
Here, $N_{\text{norm}}$ is the normalization channel yield, 
and $\varepsilon_{\text{sig}}$ and $\varepsilon_{\text{norm}}$ are the signal and normalization channel efficiencies. 
Since the branching fractions of $\D_{(s)}^+$ meson decays have been measured very precisely~\cite{PDG2024}, 
the branching fraction \mbox{$\BRof\phimm=(3.05\pm0.16)\times10^{-4}$}~\cite{LHCb-PAPER-2023-038} 
is the largest source of systematic uncertainty across all modes. 
The value of \mbox{$\BRof\dsphi=(4.50\pm 0.13)\times 10^{-2}$} is computed by combining the values reported in Ref.~\cite{PDG2024}
for the branching fractions of \mbox{$\Ds\to\phiz(\to \Kp\Km)\pip$} and \mbox{$\phiz\to \Kp\Km$} decays.
To avoid any bias, the \mumu invariant-mass regions 
within $\pm 15\mevcc$ of the known~\etaoretap mass values~\cite{PDG2024}, 
referred to as blind regions in the following, are removed from consideration in the analysis
until the selection procedure is finalized.

Muons and the pion from the $\D_{(s)}^+$ meson decay are required to satisfy
the hardware trigger selection associated with each particle species.
The \hltone step selects a two-track signature on any 
combination of the final-state particles. 
A dedicated \hlttwo selection targeting \mbox{$\D_{(s)}^+\to\pip\mumu$} is used on the \mumu channels.
For the \mbox{\pipi\mumu} modes, a combination of inclusive selections targeting multiple-track signatures is used.
Consequently, each final state is assigned its own normalization sample,
to match the \hlttwo requirement of the signal mode.

The \etaoretap candidates are formed by combining two oppositely charged tracks consistent with the muon hypothesis. 
For four-particle final states, two additional tracks consistent with the pion hypothesis are included. 
The resulting \mumu and \pipi\mumu systems are required to have an invariant mass 
within $\pm50\mevcc$ ($\pm100\mevcc$) of the known $\etaz(\etapr)$ mass, 
and, in the case of \phimm, within $\pm100\mevcc$ of the known \phiz mass~\cite{PDG2024}. 
The \DporDsp candidates are then formed by combining the \etaz, \etapr and \phiz candidates with an additional pion, 
requiring that they originate from a common vertex. 
All tracks are required not to originate from the $pp$ collision vertex (PV).
The reconstructed invariant mass is required to lie within the ranges
$[1800,1920]\mevcc$  and $[1920,2050]\mevcc$ for the \Dp and \Ds candidates, respectively.

Several physical background processes may contribute to all the decay modes.
The radiative decays \mbox{$\etaoretap\to\mumu\g$} are suppressed to negligible levels
after selection requirements on the invariant masses of the $D_{(s)}^+$ and \etaoretap mesons, 
as well as the impact parameter of the $D_{(s)}^+$ candidate. 
The decay \mbox{$\etapr\to\pipi\etaz(\to \mup\mun)$} falls inside the 
\mbox{\etapfb} signal region, 
and is vetoed through a requirement on the dimuon mass, $m_{\mup\mun}<520\mevcc$. 
In the \mbox{\etaptb} mode, \mbox{\phimm} decays populate the \mbox{$m_{\mumu}>970~\mevcc$} region 
and are kinematically indistinguishable from signal decays. 
The \etapr and \phiz peaks can, however, be separated by refitting the whole decay chain~\cite{Hulsbergen:2005pu}, 
constraining the \DporDsp candidate to its known mass~\cite{PDG2024} 
and to originate from the associated PV. 
This procedure improves the mass resolution of the resonance peaks 
and suppresses leakage into the \etapr invariant-mass region.
For consistency, it is applied to all decay modes.
Contributions from prompt \etaoretap decays are evaluated from the combinatorial background in the $\D_{(s)}^+$ spectrum,
and found to be negligible. The precision of this test is assigned as systematic uncertainty.
The remaining background contribution is combinatorial, 
which is accounted for using a boosted decision tree (BDT) classifier.

A BDT is implemented for each signal and production mode through the \textsc{XGBoost} Python package~\cite{XGBoosting}. 
It is trained using simulation as a signal proxy
and data in specific mass regions populated only by background as a background proxy. 
These regions are: \mbox{$m_{\mup\mun}\in[565,600]\mevcc$} for \mbox{\etamm} decays,
\mbox{$m_{\mup\mun}\in[880,940]\mevcc$} for~\mbox{\etaptb} decays,
and \mbox{$m_{\pip\pim\mup\mun}\in[920,940]\cup[970,1000]\mevcc$} for~\mbox{\etapfb} decays.
The variables used in the training comprise particle identification (PID) information on the muon and pion candidates, kinematics and lifetime uncertainty of the \etaoretap candidates, vertex quality, lifetime, lifetime uncertainty of the \DporDsp candidate and variables related to its pointing to its associated PV, as well as the quality of the refit of the decay chain and variables related to the isolation of the selected \DporDsp candidate with respect to other tracks in the event~\cite{thesis_isolations}.
For \mbox{\etafb} decays, the combinatorial background is negligible 
after prior selection requirements; therefore, no BDT classifier is required. 
For each production and signal decay mode, the BDT requirement is defined to optimize a given metric.
To this end, pseudoexperiments are generated 
to model the expected mass distribution of the signal in the blinded region.
The expected background in this region is extrapolated from fits to the 
data outside of it, and the expected signal is calculated using the 
branching fractions predicted by~ChPT, as provided in~\tabref{tab:status}.
The metrics, calculated using an extended maximum-likelihood fit, simultaneous to both $D$ meson production modes,
are the relative uncertainty of the signal branching fraction (for the cases of \mbox{\etamm} and \mbox{\etapfb} decays),
and the expected sensitivity to the \etapr decay, as given by Wilks' theorem~\cite{Wilks:1938dza}.

\begin{figure}[tb]
    \centering
    \begin{subfigure}{0.7\linewidth}
        \includegraphics[width=1\linewidth]{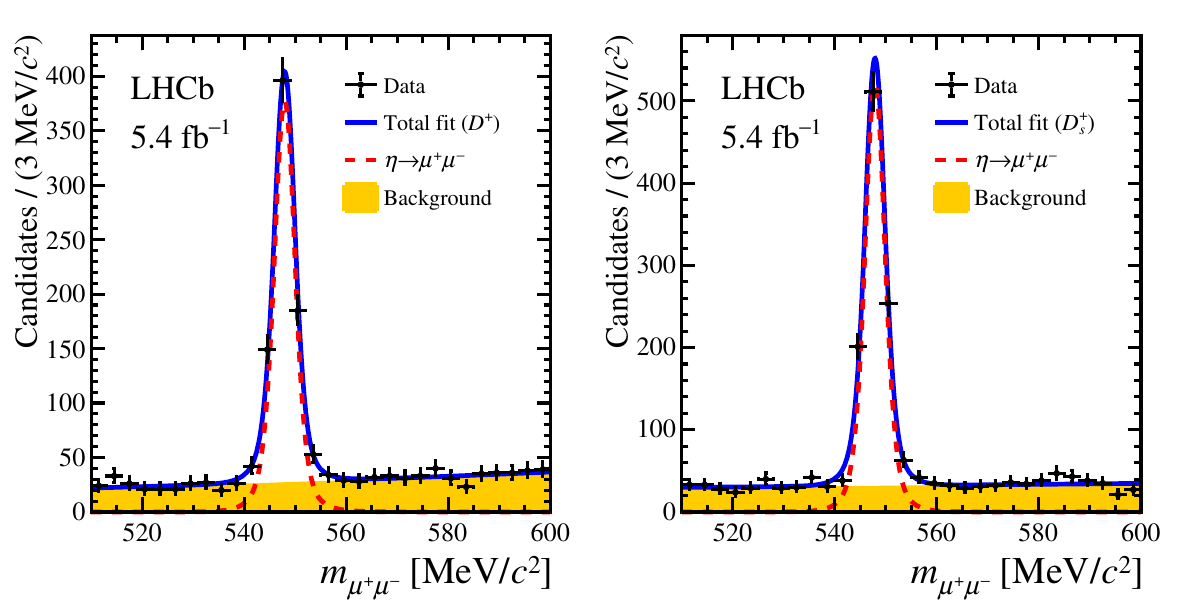}
    \end{subfigure}
    \begin{subfigure}{0.7\linewidth}
        \includegraphics[width=1\linewidth]{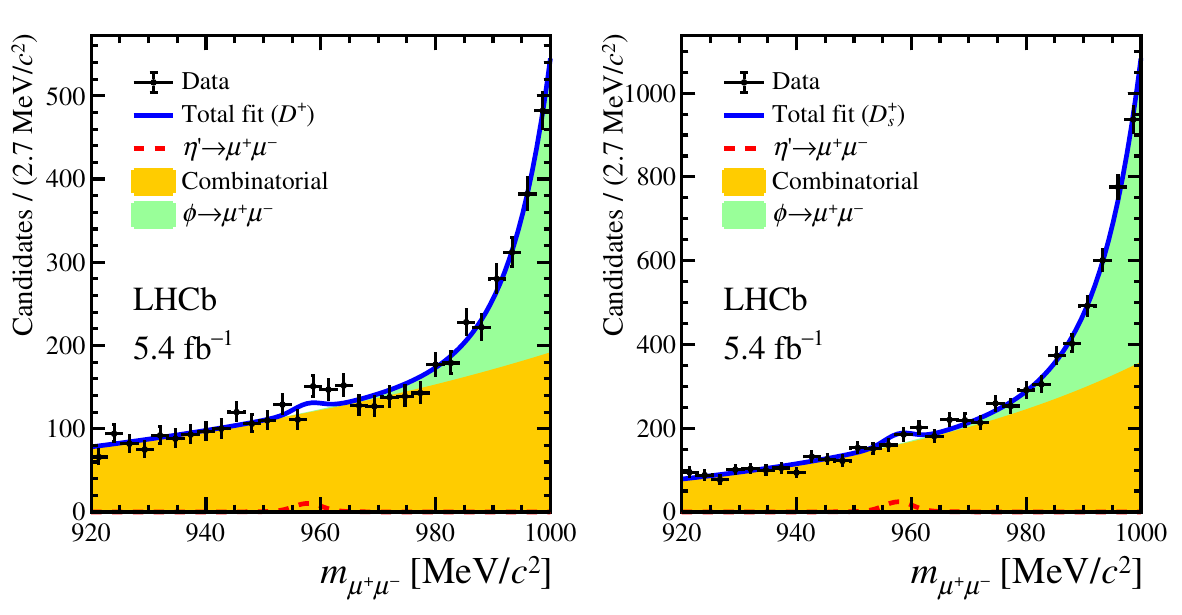}
    \end{subfigure}
    \caption{Invariant-mass distributions of the selected $\mu^+\mu^-$ candidates in the (top) $\eta$ and (bottom) $\eta^\prime$ mass regions produced in (left)  $D^+$  and (right) $D_{s}^+$ decays. 
     }
    \label{fig:finalfits_mumu}
\end{figure}

\begin{figure}
    \centering
    \begin{subfigure}{0.7\linewidth}
        \includegraphics[width=1\linewidth]{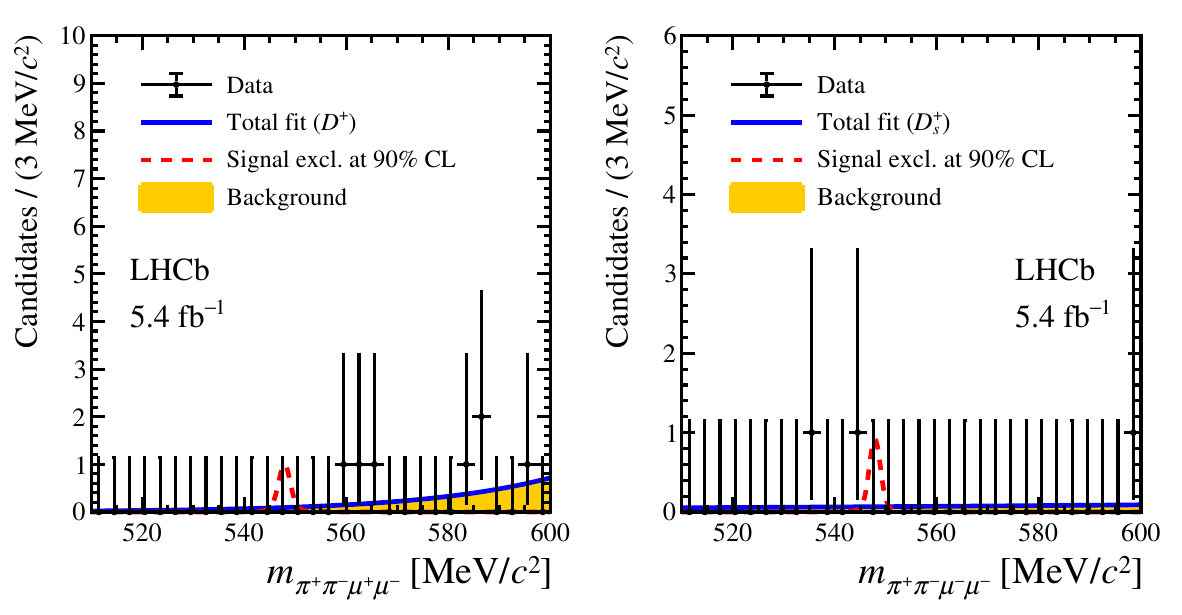}
    \end{subfigure}
    \begin{subfigure}{0.7\linewidth}
        \includegraphics[width=1\linewidth]{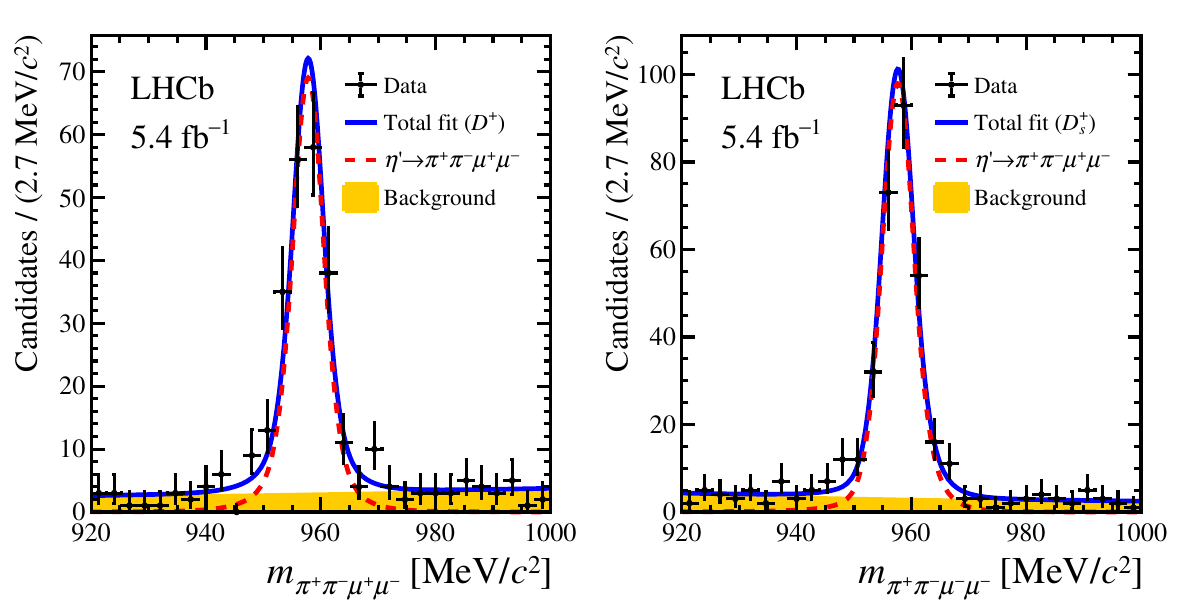}
    \end{subfigure}
    \caption{
    Invariant-mass distributions of the selected $\pi^+\pi^-\mu^+\mu^-$ candidates in the (top) $\eta$ and (bottom) $\eta^\prime$ mass regions produced in (left)  $D^+$  and (right) $D_{s}^+$ decays. 
    }
    \label{fig:finalfits_pipimumu}
\end{figure}

The selected \mbox{\mumu} and \mbox{\pipi\mumu} invariant-mass distributions are shown in \mbox{Figs.~\ref{fig:finalfits_mumu} and~\ref{fig:finalfits_pipimumu}}, where significant peaks are observed corresponding to \mbox{\etamm} and \mbox{\etapfb} decays.

To minimize systematic uncertainties related to selection efficiencies, 
the normalization channels are required to satisfy the same selection requirements
as the signal channels, except for the \mbox{$m_{\mumu}$} veto on \etapfb decays.
The BDT classifier trained on the \mbox{\mumu} signal channels is applied to the normalization modes, 
and the same selection is also required.
This effectively creates separate normalization samples for the measurements of \mbox{\etamm} and~\mbox{\etaptb} branching fractions.
The strategy is feasible due to the similar topology
and kinematic range of the final-state particles between signal and normalization.
These conditions are not satisfied in the \mbox{\pipi\mumu} channels, so no BDT classifier is applied to their normalization samples.
However, the choice of the normalization channel remains the best available option
because of the relatively large yield and signal efficiency.

The normalization channel yields are obtained from extended maximum-likelihood fits to the \phiz invariant-mass distributions, 
and are validated through separate fits to the corresponding $\D_{(s)}^+$ invariant-mass distributions.
Their differences are taken as systematic uncertainties.

Similarly, the signal channel yields are obtained from extended maximum-likelihood fits to the relevant invariant-mass distributions as detailed below.
The peaking contributions from the \etaoretap modes and the \phiz normalization channels
are modeled with a double-sided Crystal Ball probability density function (\PDF)~\cite{Skwarnicki:1986xj}, and the combinatorial background with an exponential \PDF. 
In the \mbox{\etaptb} mode, an additional exponential \PDF is included to account for the contribution from \mbox{\phimm} decays in the high-mass region.
All parameters of the signal PDFs are fixed from simulation,
after applying a correction factor to the width, 
extracted from data-simulation comparisons using the normalization channel.
Alternative signal and background models are tested by generating pseudoexperiments 
with these models and subsequently fitting them with the baseline model.
The variation in the resulting branching fraction,
which ranges between $0.01$ (for~\mbox{\etamm} and~\mbox{\etapfb}) and $0.05$ (for~\mbox{\etaptb}),
is assigned as a systematic uncertainty.
Owing to the low background and signal yield in the~\etafb mode, 
the evaluation is performed on the expected limit and found negligible.

The efficiencies are calculated using simulation for the different contributions: 
geometrical acceptance, trigger and offline selection requirements. 
\mbox{Table~\ref{tab:norm}} of the End Matter reports the ratio between normalization and signal efficiencies,
along with the normalization factors for each signal mode.
The efficiencies of the tracking, trigger, PID, and BDT selection requirements,
for both signal and normalization, are 
corrected using data calibration samples~\cite{LHCb-PUB-2014-039,LHCb-PUB-2016-021,LHCb-DP-2013-002}. 
Systematic uncertainties are assigned to these
correction factors. The tracking, trigger, and PID corrections are 
performed in kinematic bins, and the systematic uncertainty is given by the difference between binning schemes.
For the tracking and muon PID, the efficiencies are calibrated using a tag-and-probe method
on \mbox{$\jpsi\to\mumu$} candidates from data. 
The PID efficiencies of the pion from the $D_{(s)}^+$ meson are calibrated using \mbox{\dkpicf} decays.
The trigger efficiency is validated by comparing the response of the selection requirements in \dphipi data and simulation.
The efficiencies of the BDT selection requirements in the \mumu modes are validated by comparing results obtained with the nominal simulation to those from weighted samples, where the weights are derived from comparisons of the BDT input features in the \phimm control channel in data and simulation.
For the \mbox{\etapfb} production modes, the validation is performed by comparing the selection efficiency in simulation with that measured in data using the \mbox{\dsphifb} channel, which shares the same final-state topology.
Additional systematic uncertainties arise from the limited precision of the branching fractions
of \mbox{\dphipi} and \mbox{\detapi} decays, as well as from the statistical uncertainty in the efficiency due to the finite size of the simulated samples.
The breakdown of systematic uncertainties for each signal mode is provided in~\mbox{\tabref{tab:systematics}} of the End Matter.

The branching fraction of each decay mode is determined through a simultaneous extended maximum-likelihood fit to the invariant-mass distributions of both $D$-meson categories. It is treated as a common fit parameter, expressed in terms of the signal yields as in~\mbox{\refeq{eq:alpha}}, while systematic uncertainties are accounted for via Gaussian priors on the normalization factor.
Fit results are shown in \mbox{Figs.~\ref{fig:finalfits_mumu}} and~\ref{fig:finalfits_pipimumu}.
The branching fractions of \mbox{\etamm} and \mbox{\etapfb} are measured 
with a $6.8\%$ and $8.6\%$ total uncertainty, respectively
\begin{align*}
    \BRof\etamm &= (5.02\pm0.14{\stat}\pm0.18{\syst}\pm0.26\ext)\times10^{-6}\nonumber,\\
    \BRof\etapfb &= (2.73\pm0.14{\stat}\pm0.13{\syst}\pm0.14\ext)\times10^{-5}\nonumber,
\end{align*}
where the first uncertainties are statistical, the second are systematic, and the third include the contribution from \BRof\phimm.
No~\etafb signal candidates are observed, and an upper limit is established on the branching fraction of the decay
by integrating the positive side of the profile likelihood of the branching fraction:
\begin{equation}
    \BRof\etafb < 5.4\times10^{-7}{\textrm{ at 90\% CL}}\nonumber.
\end{equation}
The statistical significance of the~\mbox{\etaptb} decay, 
calculated using Wilks' theorem~\cite{Wilks:1938dza}, is $2.3\sigma$. 
The central value of the branching fraction is 
\begin{equation}
    \BRof\etaptb = (1.17\pm0.51{\stat}\pm0.06{\syst}\pm0.08\,(\text{ext}))\times10^{-7},\nonumber
\end{equation}
and the following upper limit is established:
\begin{equation}
    \BRof\etaptb < 1.9\times10^{-7} {\textrm{ at 90\% CL}}.\nonumber
\end{equation}
The likelihood profiles are included in~\figref{fig:likelihood} of the End Matter. 

The channel \mbox{\etapfb} is also used to search for BSM particles
decaying into dimuon pairs, \mbox{$\etapr\to\pipi a(\to\mumu)$}. 
For each $D_{(s)}^+$-meson production mode, a two-dimensional fit is performed
to both the \mbox{\pipi\mumu} and \mbox{\mumu} invariant-mass distributions. 
The \mbox{$a\to\mumu$} signal distribution is modeled by the product of double-sided Crystal Ball PDFs
in both the \mbox{\pipi\mumu} and \mbox{\mumu} invariant-mass distributions.
The fit model also accounts for contributions from \mbox{\etapfb} decays with no \mbox{$a\to\mumu$} contributions, and combinatorial background.
The former are described via the product of a double-sided Crystal Ball, for the \mbox{\pipi\mumu} distribution,
and an exponential function, for the \mbox{\mumu} distribution.
The latter are described by the product of exponential functions, one for each invariant-mass distribution.

Six mass values ($m_a$) for BSM resonances are explored, ranging from 250 to \mbox{500\mevcc} in steps of \mbox{50\mevcc}.
Simulation samples are generated for short-lived particles for each of the possible $m_a$ values
to fix the parameters of the double-sided Crystal Ball PDFs in the fits, 
and to determine the signal efficiencies, which are introduced in the final fit 
via the normalization factor defined in~\mbox{\refeq{eq:alpha}}.
No \mbox{$a\to\mumu$} excesses are found and upper limits
for each value of $m_a$ are reported in~\mbox{\tabref{tab:limits_dimuon}}.

\begin{table}
    \centering
    \caption{
    Upper limits on the branching fraction $\BR(\etapr\to\pipi a(\to\mumu))$ for different BSM masses $m_a$.
    }
    \label{tab:limits_dimuon}

    \begin{tabular}{lcccccc}
    \hline
    $m_a\,[\mevcc]$  & 250 & 300 & 350 & 400 & 450 & 500 \\
    \hline
    \BF($\times10^{-7})$ at 90\% CL & $<10.7$ & $<5.5$ & $<4.3$ & $<2.3$ & $<1.3$ & $<1.0$ \\
    \hline
    \end{tabular}
\end{table}

In summary, this Letter presents the first upper limit on the branching fraction~\mbox{\BRof\etappdgtb},
as well as an updated value of the branching fraction~\mbox{\BRof\etamm}, with a total uncertainty of $6.8\%$, which improves upon the current precision of $15\%$~\cite{PDG2024}.
It also provides a branching fraction measurement for~\mbox{\etappdgfb}, with a total uncertainty of $8.6\%$, and upper limit on $\etafb$.

The precision of the branching fractions \mbox{\BRof\etamm} and \mbox{\BRof\etappdgfb} is systematically limited by the uncertainty on that of \mbox{\phimm} decay.

\section*{Acknowledgements}
%
%
\noindent We gratefully acknowledge Maxim Pospelov for encouraging this type of searches at LHCb.
We express our gratitude to our colleagues in the CERN
accelerator departments for the excellent performance of the LHC. We
thank the technical and administrative staff at the LHCb
institutes.
We acknowledge support from CERN and from the national agencies:
ARC (Australia);
CAPES, CNPq, FAPERJ and FINEP (Brazil); 
MOST and NSFC (China); 
CNRS/IN2P3 and CEA (France);  
BMFTR, DFG and MPG (Germany);
NKFIH (Hungary);              
INFN (Italy); 
NWO (Netherlands); 
MNiSW and NCN (Poland); 
MEC/IFA (Romania); 
MICIU and AEI (Spain);
SNSF and SER (Switzerland); 
NASU (Ukraine); 
STFC (United Kingdom); 
DOE NP and NSF (USA).
We acknowledge the computing resources that are provided by ARDC (Australia), 
CBPF (Brazil),
CERN, 
IHEP and LZU (China),
IN2P3 (France), 
KIT and DESY (Germany), 
INFN (Italy), 
SURF (Netherlands),
Polish WLCG (Poland),
IFIN-HH (Romania), 
PIC (Spain), CSCS (Switzerland), 
GridPP (United Kingdom),
and NSF (USA).  
We are indebted to the communities behind the multiple open-source
software packages on which we depend.
Individual groups or members have received support from
RTP (Australia), 
FWO Odysseus grant G0ASD25N (Belgium), 
Key Research Program of Frontier Sciences of CAS, CAS PIFI, CAS CCEPP (China); 
Minciencias (Colombia);
EPLANET, Marie Sk\l{}odowska-Curie Actions, ERC and NextGenerationEU (European Union);
A*MIDEX, ANR, IPhU and Labex P2IO, and R\'{e}gion Auvergne-Rh\^{o}ne-Alpes (France);
Alexander-von-Humboldt Foundation (Germany);
ICSC (Italy); 
Severo Ochoa and Mar\'ia de Maeztu Units of Excellence, GVA, XuntaGal, GENCAT, InTalent-Inditex and Prog.~Atracci\'on Talento CM (Spain);
the Leverhulme Trust, the Royal Society and UKRI (United Kingdom).

\clearpage

\section*{End Matter}
\label{sec:endmatter}

The systematic uncertainties for each mode are listed in~\tabref{tab:systematics}.
The nonprompt contribution includes the comparison between the signal yields
in the normalization channel between the $\D_{(s)}^+$ and the \phiz, 
as well as the precision test against prompt decays described in the main text.

\tabref{tab:norm} provides the values involved in the determination of the normalization factors in Eq.~\eqref{eq:alpha}.

\begin{figure}
    \centering
    \includegraphics[width=0.4\linewidth]{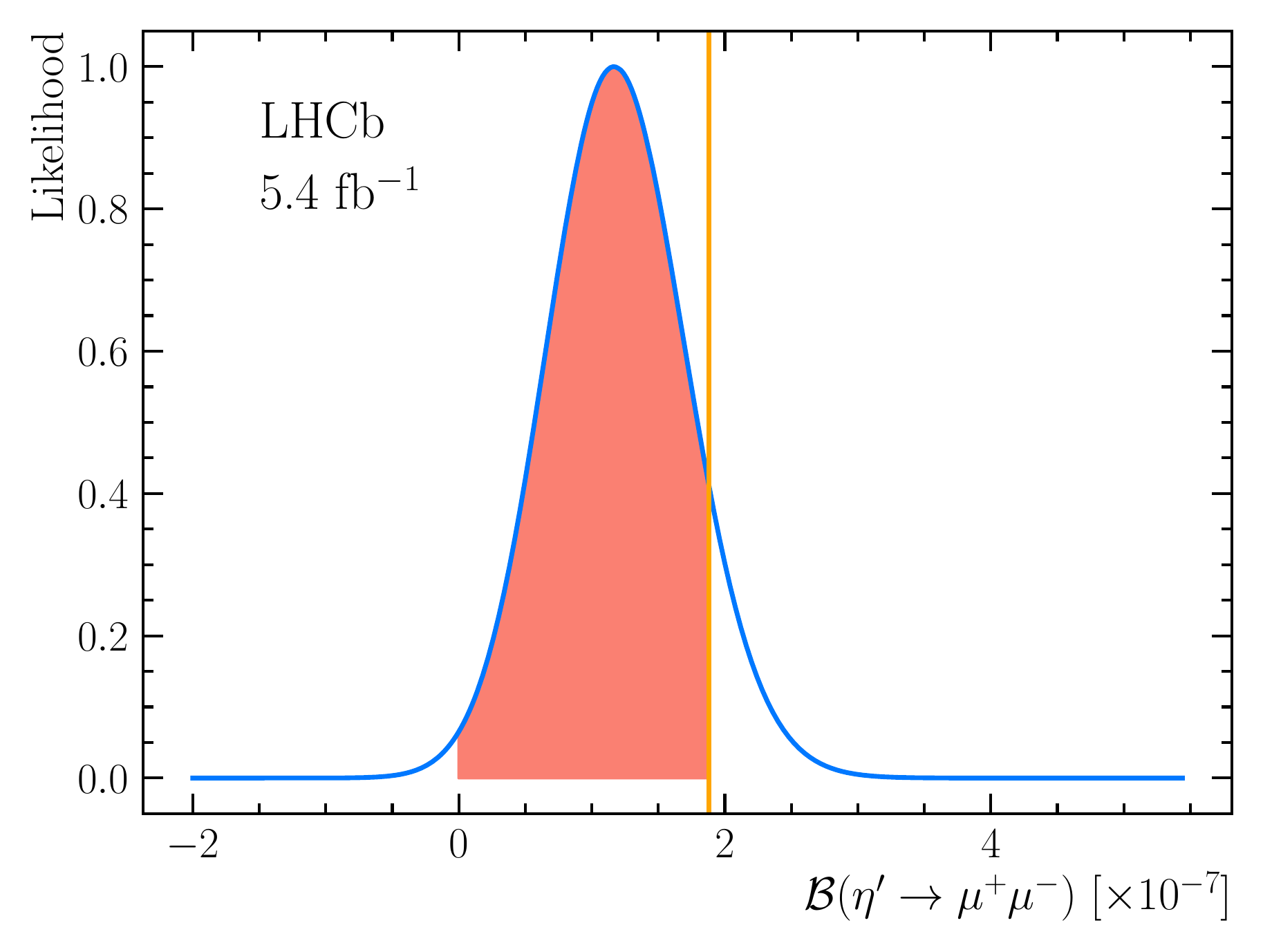}
    \includegraphics[width=0.4\linewidth]{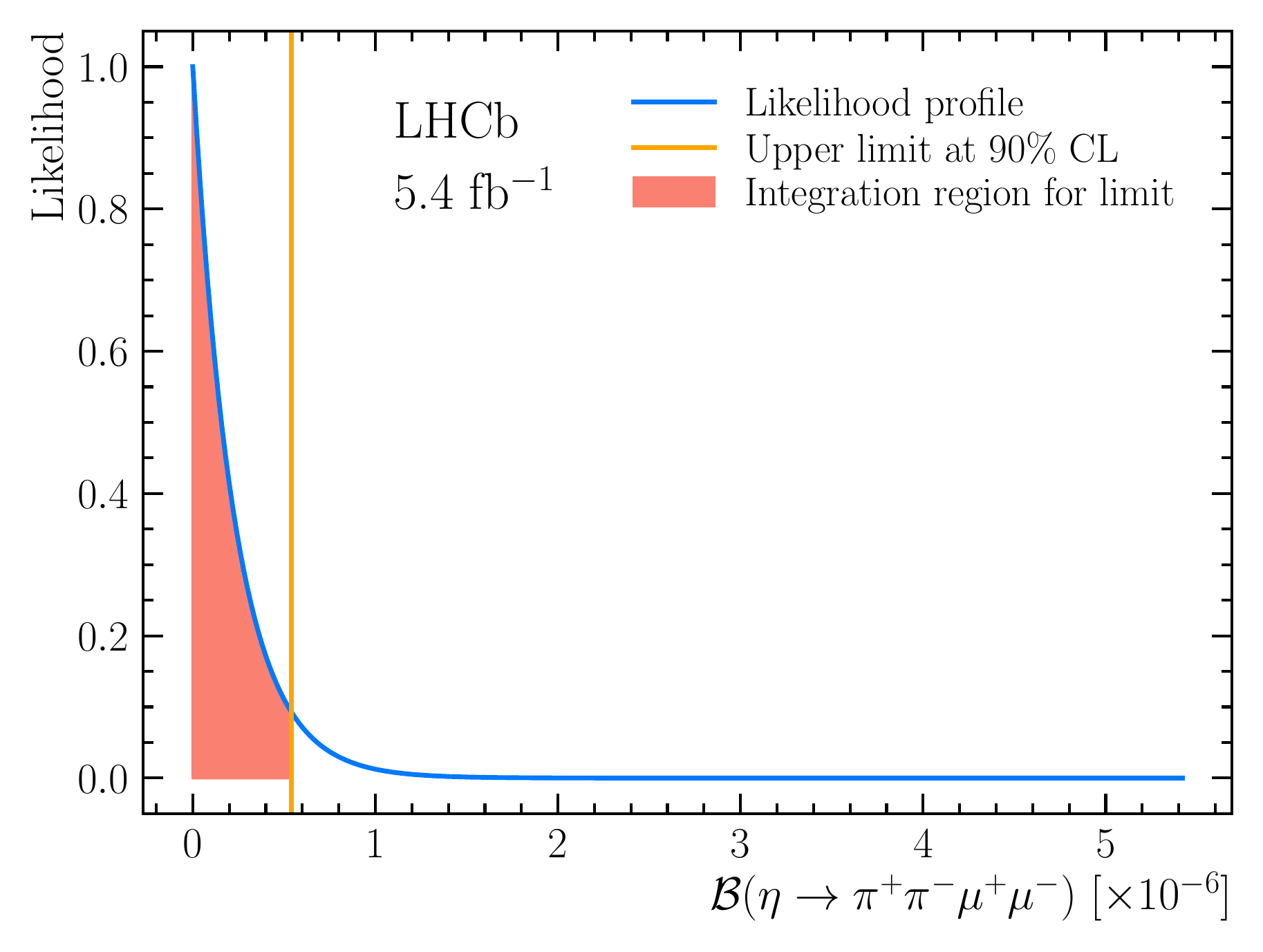}
    \caption{Likelihood profiles of (left) \mbox{$\eta^\prime\to\mu^+\mu^-$} and (right) \mbox{$\eta\to\pi^+\pi^-\mu^+\mu^-$} branching ratios. 
    They are used to calculate the upper limits 
    by integrating the positive side of the curves up to 90\% of its area.
    The \mbox{$\eta\to\pi^+\pi^-\mu^+\mu^-$} likelihood profile is displayed beginning at the point where the fit converges, and the covariance matrix can be accurately determined.
    }
    \label{fig:likelihood}
\end{figure}

The likelihood profiles for both~\mbox{\etafb} and~\mbox{\etaptb} decays are shown in~\figref{fig:likelihood}.

\begin{table}[h]
    \centering
    \caption{Sources of relative systematic uncertainties on the branching fraction measurements, in percent. 
    No BDT is trained on the mode \etafb, therefore, no systematic is assigned. 
    }
    \label{tab:systematics}
    \begin{tabular}{lcc|cc|C{1.5cm}C{1.5cm}|C{1.5cm}C{1.5cm}}
    \hline
       & \multicolumn{2}{c|}{\etamm} & \multicolumn{2}{c|}{\etaptb} & \multicolumn{2}{c|}{\etafb} & \multicolumn{2}{c}{\etapfb}  \\
    \hline
    Source & \Dp & \Ds & \Dp & \Ds & \Dp & \Ds & \Dp & \Ds \\
    \hline
    $\frac{\BRof\dphipi}{\BRof\detapi}$  & 3.4 & 3.4 &  4.5 &  3.5 &   3.4  &   3.4  &   4.5    &  3.5   \\
    Trigger     & 0.9 & 0.4 &  0.2 &  0.3 &   5.3  &   1.7  &   2.7   &   1.7 \\
    BDT         & 0.6 & 1.4 &  4.3 &  1.4 &   ---  &   ---  &   0.9   &   1.3   \\
    PID         & 0.2 & 0.2 &  0.1 &  0.2 &   1.7  &   1.8  &   1.0   &   1.1  \\
    Tracking    &  2.2 & 2.2 &  2.2 &  2.2 &   3.3  &   3.2  &   3.2   &   3.1 \\
    Simulation size    & 0.4 & 0.5 & 0.8 & 0.9 & 1.1 & 1.9 & 1.5 & 2.9 \\
    Nonprompt    & 3.0 & 2.3 &  3.0 &  2.3 &   3.0  &   2.3  &   3.6   &   2.3 \\
    \hline
    Total       & 7.3 & 7.1 &  8.9 &  7.2 &   9.5  &   7.9  &   9.0   &   8.2          \\
    \hline
    \end{tabular}
\end{table}

\begin{table}
\centering
\caption{Quantities involved in the computation of the normalization factor of each signal channel. 
The value $\BRof\phimm = (3.05\pm0.16)\times10^{-4}$~\cite{LHCb-PAPER-2023-038} has been factored into the final row.
Systematics are also taken into account in the uncertainty of \Palpha.
}\label{tab:norm}
\begin{tabular}{lcc|cc}
\hline
& \multicolumn{2}{c|}{\etamm} & \multicolumn{2}{c}{\etaptb}  \\
\hline
Source & \Dp & \Ds & \Dp & \Ds \\
\hline
$\frac{\BRof\dphipi}{\BRof\detapi}$ & 
$1.512\pm0.052$ & $2.663\pm0.091$ & $1.147\pm0.052$ & $1.139\pm0.040$  \\
$\varepsilon_{\text{norm}} / \varepsilon_{\text{sig}}$ &
$1.427\pm0.011$ & $1.384\pm0.013$ & $1.1039\pm0.0039$ & $1.0941\pm0.0052$ \\
$N_{\text{norm}}$ ($\times10^{3}$) & 
$90.59\pm0.31\phz$ & $188.42\pm0.46\phz\phz$ & $87.22\pm0.30\phz$ & $179.81\pm0.44\phz\phz$  \\
\Palpha ($\times10^{-9}$) & 
$7.25\pm0.55$ & $5.30\pm0.38$ & $4.45\pm0.47$ & $1.85\pm0.16$  \\ 
\hline
\hline
& \multicolumn{2}{c|}{\etafb} & \multicolumn{2}{c}{\etapfb} \\
\hline
Source & \Dp & \Ds & \Dp & \Ds\\
\hline
$\frac{\BRof\dphipi}{\BRof\detapi}$ &  
$1.512\pm0.052$ & $2.663\pm0.091$ & $1.147\pm0.052$ & $1.139\pm0.040$ \\
$\varepsilon_{\text{norm}} / \varepsilon_{\text{sig}}$ &
$56.66\pm0.87\phz$ & $65.20\pm0.19\phz$ & $26.01\pm0.28\phz$ & $32.09\pm0.60\phz$ \\
$N_{\text{norm}}$ ($\times10^{3}$) &  
$77.02\pm0.36\phz$ & $125.52\pm0.41\phz\phz$ & $77.02\pm0.36\phz$ & $125.52\pm0.41\phz\phz$ \\
\Palpha ($\times10^{-9}$) &  
$432\pm41\phz$ & $418\pm35\phz$ & $149\pm13\phz$ & $91.0\pm7.2\phz$ \\ 
\hline
\end{tabular}
\end{table}

\clearpage

\addcontentsline{toc}{section}{References}
\bibliographystyle{LHCb}
\bibliography{main_ana_V2,main,standard,LHCb-PAPER,LHCb-CONF,LHCb-DP, LHCb-TDR}

\newpage
\centerline
{\large\bf LHCb collaboration}
\begin
{flushleft}
\small
R.~Aaij$^{38}$\lhcborcid{0000-0003-0533-1952},
A.S.W.~Abdelmotteleb$^{58}$\lhcborcid{0000-0001-7905-0542},
C.~Abellan~Beteta$^{52}$\lhcborcid{0009-0009-0869-6798},
F.~Abudin\'en$^{58}$\lhcborcid{0000-0002-6737-3528},
T.~Ackernley$^{62}$\lhcborcid{0000-0002-5951-3498},
A.A.~Adefisoye$^{70}$\lhcborcid{0000-0003-2448-1550},
B.~Adeva$^{48}$\lhcborcid{0000-0001-9756-3712},
M.~Adinolfi$^{56}$\lhcborcid{0000-0002-1326-1264},
P.~Adlarson$^{86}$\lhcborcid{0000-0001-6280-3851},
C.~Agapopoulou$^{14}$\lhcborcid{0000-0002-2368-0147}
% ...

{\footnotesize \it

% ...
$^{14}$Universit{\'e} Paris-Saclay, CNRS/IN2P3, IJCLab, Orsay, France\\
% ...
$^{38}$Nikhef National Institute for Subatomic Physics, Amsterdam, Netherlands\\
% ...
$^{48}$Instituto Galego de F{\'\i}sica de Altas Enerx{\'\i}as (IGFAE), Universidade de Santiago de Compostela, Santiago de Compostela, Spain\\
% ...
$^{52}$Physik-Institut, Universit{\"a}t Z{\"u}rich, Z{\"u}rich, Switzerland\\
% ...
$^{56}$H.H. Wills Physics Laboratory, University of Bristol, Bristol, United Kingdom\\
% ...
$^{58}$Department of Physics, University of Warwick, Coventry, United Kingdom\\
% ...
$^{61}$School of Physics and Astronomy, University of Glasgow, Glasgow, United Kingdom\\
$^{62}$Oliver Lodge Laboratory, University of Liverpool, Liverpool, United Kingdom\\
% ...
$^{70}$Syracuse University, Syracuse, NY, United States\\
% ...
$^{86}$Department of Physics and Astronomy, Uppsala University, Uppsala, Sweden, associated to $^{61}$\\
% ...
}
\end{flushleft}
\end{document}